\documentstyle[11pt]{article}
\input epsf
\newcommand{\Rsl}{{\not \! \!{R}}}

\newcommand{\sun}{\Delta m^2_{\rm solar}}
\newcommand{\atm}{\Delta m^2_{\rm atm}}
\newcommand{\lsnd}{\Delta m^2_{\rm LSND}}
\def\bea{\begin{eqnarray}}   
\def\eea{\end{eqnarray}}

\begin{document}
\vspace*{-1in}
\renewcommand{\thefootnote}{\fnsymbol{footnote}}
\begin{flushright}
LPT Orsay/00-58 \\
SINP/TNP/00-14\\
\texttt{hep-ph/0007016} 
\end{flushright}
\vskip 5pt
\begin{center}
{\Large {\bf Can R-parity violation explain the LSND data as well?}}
\vskip 25pt
{\bf Asmaa Abada $^{1}$}
~and~  
{\bf Gautam Bhattacharyya $^{2}$}
\vskip 10pt  
$^1${\it Laboratoire de Physique Th\'eorique, 
Universit\'e de Paris XI, B\^atiment 210, \\ 91405 Orsay Cedex,
France} \\
$^2${\it Saha Institute of Nuclear Physics, 1/AF Bidhan Nagar, 
Calcutta 700064, India}
\vskip 20pt
{\bf Abstract}
\end{center}

\begin{quotation}
  {\noindent\small The recent Super-Kamiokande data now admit only one
    type of mass hierarchy in a framework with three active and one
    sterile neutrinos.  We show that neutrino masses and mixings
    generated by R-parity-violating couplings, with values within their
    experimental upper limits, are capable of reproducing this
    hierarchy, explaining all neutrino data particularly after including
    the LSND results.

\vskip 5pt
\noindent
PACS number(s):~12.60.Jv, 14.60.Pq, 11.30.Fs
}

\end{quotation}

\vskip 15pt  

\setcounter{footnote}{0}
\renewcommand{\thefootnote}{\arabic{footnote}}

The vacuum oscillation interpretation of solar neutrino data requires
$\sun \sim 10^{-10} ~{\rm eV}^2$, while the matter enhanced MSW
solution prefers the range $\sun \sim (10^{-5} - 10^{-4})~ {\rm eV}^2$
\cite{solar}. The atmospheric neutrino anomaly can be explained by
$\atm \sim (2 - 5) \times 10^{-3} ~{\rm eV}^2$ (with $\sin^2
2\theta_{\rm{atm}}> 0.88$) \cite{superk,newsk}.  In the standard three
neutrino framework, which offers two independent mass differences, it
is possible to conceive of mass hierarchies which can explain both
solar and atmospheric results. If in addition the LSND data, carrying
a positive indication of $\nu_\mu \leftrightarrow \nu_e$ oscillation
with $\lsnd \sim (0.3 - 1) ~{\rm eV}^2$ \cite{lsnd}, are also sought
to be simultaneously explained, then one has to expand the standard
three neutrino scenario as it falls short of one independent mass
difference. In fact, within this three neutrino framework one can fit
any two of the three $\Delta m^2$ mentioned above. Under the
circumstances, one often leaves the LSND data out of consideration
pending further confirmation from the MiniBooNE \cite{miniboone} at
FNAL or MINOS long baseline \cite{minos} experiments, since KARMEN
\cite{karmen,joint} does neither confirm nor exclude the LSND results.
On the other hand, if one includes the LSND results to be explained
together with the solar and atmospheric data, the minimal extension of
the standard scenario that needs to be done is to add one sterile
neutrino ($\nu_s$) \cite{pont,sterile} to the list of three active
states. The resulting four neutrino picture can explain all the data
\cite{4nudata}.

What are the possible choices of mass hierarchies among these four
neutrinos? The data suggest that the choices are very limited. First,
we devide them into two types, I and II. In type I, there are three
almost degenerate states explaining the solar and atmospheric results,
with a fourth state separated by a large gap. This hypothesis cannot
explain the LSND results, since the LSND data require that the
separated fourth state will have to be either $\nu_e$ or $\nu_\mu$,
but each of them will have to be closely spaced with a third state to
explain the solar and atmospheric results. If one disregards the LSND
results till further confirmation, then of course type I scenario is
allowed with the isolated state either $\nu_s$ or $\nu_\tau$. In type
II, there are two pairs of approximately degenerate states separated
by a large gap. The mixing between the two pairs is very small.  This
scenario can explain all the data, or in other words, the large gap
could be the LSND gap.  Two cases may arise in this framework. In type
IIa scenario, ($\nu_\tau$-$\nu_e$) form one pair which explains the
solar neutrino data, while ($\nu_\mu$-$\nu_s$) form the other pair
maximally mixed to explain the atmospheric anomaly. In type IIb,
($\nu_\tau$-$\nu_\mu$) form the pair that explains the atmospheric
anomaly, while ($\nu_s$-$\nu_e$) pair explains the solar data
\footnote{$\nu_e \leftrightarrow \nu_s$ solar neutrino oscillation
will be tested by the SNO experiment soon \cite{sno}.}. The most
recent Super-Kamiokande (SK) data rule out the $\nu_\mu
\leftrightarrow \nu_s$ interpretation of atmospheric neutrino anomaly
at 99\% CL \cite{newsk}, thus strongly disfavouring the type IIa
scenario. We are then left with type IIb as the only surviving option,
shown in Fig.~1\footnote{This framework is consistent with the
Big-Bang Nucleosynthesis constraints on sterile - active mixings
\cite{dolgov}.}. It should be noted that the oscillation data cannot
discriminate between cases that occur by interchanging either the
members within a given pair or the relative spacing of the two
pairs. These cases are not separately shown.

Under the circumstances, it is a timely exercise to identify those
scenarios which reproduce the type IIb spectrum. Do the
R-parity-violating ($\Rsl$) supersymmetric models \cite{rpar} fall in
that category? In this note we seek to find an answer to this
question.  Defined in terms of lepton number ($L$), baryon number
($B$) and spin ($S$) of the particle as $R = (-1)^{3B+L+2S}$, this
discrete parity is +1 for all SM particles and -1 for their
superpartners. Since neither $L$- nor $B$-conservation is ensured by
gauge invariance or any such fundamental principles, R-parity is an
{\em ad hoc} symmetry put in by hand. Although there is no
experimental confirmation yet in favour of non-vanishing $\Rsl$
interactions, the neutrino oscillation data are somehow suggestive
that it would be premature to abandon those couplings, as the origin
of neutrino masses and mixings could be traced to some of these
non-vanishing $L$-violating couplings.  In order not to allow rapid
proton decay we do not switch on $L$ and $B$ violations
simultaneously. The following $L$-violating terms in the
superpotential are then allowed: (a) $\lambda_{ijk} L_iL_jE^c_k$, (b)
$\lambda'_{ijk} L_iQ_jD^c_k$, and (c) $\mu_i L_i H_u$. In these
expressions, $L_i$ and $Q_i$ are SU(2)-doublet lepton and quark
superfields, $E^c_i$ and $D^c_i$ are SU(2)-singlet charged lepton and
down quark superfields, and $H_u$ is the Higgs superfield that gives
masses to up-type quarks. Stringent experimental constraints on these
{\em a priori} independent parameters exist in the literature
\cite{review}.

Attempts to fit the observed neutrino data by masses and mixing angles
generated by $\Rsl$ couplings have been done in the past, both in the
context of bilinear \cite{bilinear} as well as trilinear
\cite{trilinear} $L$-violating parameters. Most of the analyses have
been carried out in the three-neutrino picture. A few remarks on those
analyses are in order. Most of them discarded the LSND data as
something not so reliable, and confined the discussions to the
possibility of fitting only solar and atmospheric data. Adhikari and
Omanovic \cite{trilinear}, on the other hand, tried to fit solar,
atmospheric ({\em preliminary} SK) and LSND all together in a
three-neutrino picture and that too considering only trilinear $\Rsl$
couplings. But they did not consider the SK zenith angle dependence
and also assumed an energy independent solar neutrino solution by
ignoring the Chlorine data of the Homestake mine experiment. In fact,
if one takes all {\em present} data into consideration, it is not
enough just to add bilinear $\Rsl$ terms to their analysis -- the data
compel one to go for a four-neutrino picture.  Ref.~\cite{chun} deals
with the simultaneous presence of bilinear and trilinear
parameters. Although the emphasis in ref.~\cite{chun} is mostly on
finding an explanation for the solar and atmospheric neutrino data in
a three-neutrino framework, a qualitative discussion on how to
simultaneously explain the LSND data by admitting a fourth sterile
state has also been presented.

In the present work, we perform a numerical study to examine whether
R-parity violation, with bilinear and trilinear parameters together,
can reproduce the type IIb spectrum. Even though the data allow an
interchange of the location of the two pairs, we work in a situation
where ($\nu_\mu$-$\nu_\tau$) form the heavier pair\footnote{This
hierarchy is stable under radiative corrections \cite{IbNa}.}. We assume
that the masses of all active neutrinos are generated by R-parity
violation. Since the mixing between the two pairs is small, we can,
for all practical purposes, focus on the heavier pair and work in the
$\nu_\mu$--$\nu_\tau$ subspace, always assuming that the parameters
$\lambda'_{1jk}$ and $\mu_1$, responsible for the $\nu_e$ mass
generation, are much smaller. We also assume that the sterile state
receives mass of the order of the $\nu_e$ mass from some other
source. More specifically, we concentrate on those parameters which
generate the masses and mixing angles in the above ($2\times 2$)
subsector, keeping in mind that each of the two absolute masses should
be of the order of the LSND gap. In other words, we parametrize the
$\nu_\mu$--$\nu_\tau$ mass matrix in terms of the $\Rsl$ couplings,
vary them within physical ranges, and then observe whether there exist
solutions that simultaneously satisfy the following experimental
constraints:
\bea
\left\{
 \begin{array}{ll}
 \Delta m^2_{\rm LSND} \sim m_4^2 \sim m_3^2 \sim (0.3 - 1) 
 ~{\rm eV}^2,  \\
  \Delta m^2_{\rm atm}=m_4^2-m_3^2
 \sim (2 - 5) \times 10^{-3} ~{\rm eV}^2, \\
 \sin^2 2\theta_{\rm{atm}}\equiv \sin^2 2\theta_{{34}} >  0.88. 
 \end{array}
 \right.
 \label{constraints}
 \eea

 Now we turn our attention to the analytic expressions of the neutrino
 masses induced by bilinear and trilinear couplings. We write the mass
 matrix as ${\cal{M}_{\nu}}={\cal{M}}^{\rm{tree}}_{\nu}
 +{\cal{M}}^{\rm{loop}}_{\nu}$. The bilinear couplings contribute to
 the tree mass \cite{bilinear}. In a basis where there are no
 sneutrino vacuum expectation values (VEVs) \cite{basis}, this can be
 expressed as\footnote{In principle, one can rotate away the bilinear
 $\mu_i$ terms from the superpotential, but still having the $\Delta L
 =2$ effects via the presence of sneutrino VEVs after the minimization
 of the scalar potential. One can as well work in a framework where
 there are no sneutrino VEVs but the $\mu_i$ parameters are present in
 the theory \cite{basis}. We work in the latter basis.  We emphasize
 that going from one basis to another, the parametrization would
 change, but the general conclusion we draw at the end remains
 unaffected. For a discussion on the basis-independent
 parametrizations of R-parity violation, see
 refs.~\cite{basis-independent}.}  \bea
 {\cal{M}}^{\rm{tree}}_{\nu_{ii'}} = g_{2}^{2}{(M_{1} +
 \tan^2\theta_{W} M_{2})\over 4 \det M}\mu_{i}\mu_{i'} v_{d}^{2}
 \equiv \alpha \mu_{i}\mu_{i'}~~~~~(v_d = \langle H_d^0 \rangle) \ ,
\label{treeelem}
\eea
where $M_{1,2}$ are the gaugino masses, and $\det M$ is the
determinant of the $(4\times 4)$ neutralino mass matrix in the
R-parity-conserving case.  Considering only the $\lambda'$ couplings,
the one-loop squark-mediated contribution to the neutrino mass, in the
same basis as before, can be written as \cite{trilinear}
\begin{equation} 
{\cal{M}}^{\rm{loop}}_{\nu_{ii'}} \simeq {{N_c \lambda'_{ijk} 
\lambda'_{i'kj}}
\over{16\pi^2}} m_{d_j} m_{d_k}
\left[\frac{f(m^2_{d_j}/m^2_{\tilde{d}_k})} {m_{\tilde{d}_k}} +
\frac{f(m^2_{d_k}/m^2_{\tilde{d}_j})} {m_{\tilde{d}_j}}\right],
\label{mass}
\end{equation}     
where $f(x) = (x\ln x-x+1)/(x-1)^2$. Here, $m_{d_i}$ is the down quark
mass of the $i$th generation, $m_{\tilde{d}_i}$ is an average of
$\tilde{d}_{Li}$ and $\tilde{d}_{Ri}$ squark masses, and $N_c = 3$ is
the colour factor. While writing Eq.~(\ref{mass}), we assumed that the
left-right squark mixing terms are family-diagonal and are
proportional to the corresponding quark masses, i.e., $m^2_{\rm LR}
(i) = m_{d_i} m_{\tilde{d}_i}$. The expression of the
$\lambda$-induced slepton-mediated contributions to the neutrino mass
is similar to Eq.~(\ref{mass}), and we do not display it
here\footnote{We neglect the one-loop diagrams induced by the product
  of bilinear and trilinear couplings \cite{others}.}.

In the basis $(\nu_\mu,\ \nu_\tau)$, the mass matrix can be 
 parametrized as 
\bea {\cal{M}}_{\nu}= \left(
 \begin{array}{cc}
K\lambda_2^2 + \alpha \mu_2^2 & 
K\lambda_2 \lambda_3+ \alpha \mu_2 \mu_3 \\
K\lambda_2\lambda_3+ \alpha \mu_2 \mu_3  &  
K\lambda_3^2+ \alpha \mu_3^2 
 \end{array}
 \right), 
\label{toy1}
\eea 
where $\alpha$ is given by Eq.~(\ref{treeelem}), $\lambda_{2,3}$
are two generic trilinear couplings, and $K$ captures the loop
factors that enter into Eq.~(\ref{mass}).  
  
We observe that the determinant of the mass matrix with only bilinear or
with only trilinear couplings is identically zero\footnote{If we have
more than one combination of trilinear couplings, then although the
determinant of the mass matrix with trilinear couplings alone will not
be zero, still much more fine-tuning may be necessary in order to arrive
at any solution with only trilinear couplings.}. But this yields a big
hierarchy between $m_3$ and $m_4$, contrary to the requirement that
these two states should be approximately degenerate. Only after taking
the bilinear and trilinear parameters together we obtain solutions that
pass the test in Eq.~(\ref{constraints}). Actually we need at least four
independent input parameters in the mass matrix in order to fit the
data. In this case, they are the two trilinear couplings ($\lambda_2$,
$\lambda_3$) and the two bilinear mass parameters ($\mu_2$, $\mu_3$).
Notice that these four parameters as well as $\alpha$ can take either
sign, while $K$ is always positive.  We point out that the tree and loop
contributions may have the same or opposite signs, and this relative
sign plays a crucial role in deciding which combinations of parameters
are allowed by the data.  After diagonalizing the mass matrix in
Eq.~(\ref{toy1}), we demand that the eigenvalues $m_{3,4}$ and the
mixing angle $\theta_{34}=\theta_{\rm atm}$ satisfy
Eq.~(\ref{constraints}). In Figs.~2a and 2b we have displayed only a
part of the solutions by plotting the acceptable mass spectrum as a
function of some allowed input parameters, just to demonstrate that the
mechanism of R-parity violation works as a viable
explanation\footnote{In principle, the allowed spectrum should have been
displayed as a five-dimensional plot where all the four input parameters
are varying. Mostly for the purpose of a simplified presentation, merely
to point out that there indeed exist solutions satisfying
Eq.~(\ref{constraints}), we plotted the spectrum in two dimensions in
Figs.~2a and 2b. In each plot, the `other three' parameters are also
varying.}. The above parametrization is rather general as one can apply
it for any $\lambda'$ (or $\lambda$, for that matter) couplings
irrespective of the second and third generation indices.  Also at this
stage one need not specify the squark (or slepton) and gaugino masses as
they are absorbed in $K$ and $\alpha$ respectively. We observe that the
`filter' of Eq.~(\ref{constraints}) prefers a negative $\alpha$, which
implies that one or more gaugino mass parameters could be negative (see
Eq.~(\ref{treeelem})). We stress though that we do get some solutions
with positive values of $\alpha$ as well.

Now assuming, as an illustrative example, that $\lambda'_{233}$ and
$\lambda'_{233}$ are the only dominant trilinear couplings, the factor
$K$ turns out to be $K \sim N_c m_b^2/8\pi^2 m_{\tilde{b}}$. Taking
the squarks and gauginos to be approximately at the $300$ GeV scale,
one obtains $\alpha \sim 2.~10^{-4}~ {\rm GeV}^{-1}$, and $K \sim
1.~10^{-3}$ GeV. The minimum and maximum values of the input
parameters that pass this test turn out to be
\bea 
&\lambda'_{233} \sim
\lambda'_{333} : [-1.3\times 10^{-3}~ , ~ 1.3\times 10^{-3}];
\nonumber \\
& \mu_{2,3} ~({\rm GeV})~ : [-5.0\times 10^{-3}~,~ 1.0\times 10^{-3}].
\eea
These values are consistent with the existing constraints on the above
parameters \cite{review,fcnc}.

To conclude: If R-parity violation has to explain all neutrino data,
it is essential to have both trilinear and bilinear $\Rsl$ terms in
addition to having a sterile neutrino in the model. Then it is
possible to generate maximally mixed $\nu_\mu$ and $\nu_\tau$ with
their absolute masses in the eV range and mass-squared difference in
the milli-${\rm eV}^2$ range, as required by the data. The quoted
ranges of the $\Rsl$ parameters within which we obtain solutions are
based on certain simple-minded but plausible approximations made for
the ease of presentation. The bottom line of our analysis is that we
provide an affirmative answer to the question we have asked in the
title.

\noindent 
{\em \underline{Note added}}: While we were finishing this note, we
became aware of a {\em preliminary} SK solar neutrino analysis update
\cite{Totsuka} disfavouring a pure $\nu_e \leftrightarrow \nu_s$ solar
neutrino oscillation at 95\% CL. First, this is only a 2-$\sigma$
result which is not enough to exclude a model. Second, it has been
claimed that a rate $+$ spectrum combined analysis in a full
four-flavour scenario exhibits an allowed zone \cite{concha}.  So the
fate of the sterile state may not be that dwindling, and we must wait
till the SNO experiment tests this option.

\vskip 10pt
\noindent 
Both authors thank the CERN Theory Division, where this work has been
done, for its warm hospitality. They also thank H. Dreiner and
A. Raychaudhuri for illuminating comments on the manuscript. GB
also thanks LPT, Orsay, for its hospitality, and S. Choubey and
S. Goswami for clarifying remarks on neutrino data.  

\newpage   

\newpage 

\begin{figure}[!ht]
\vskip 10.2in\relax\noindent\hskip -1in\relax
{\includegraphics{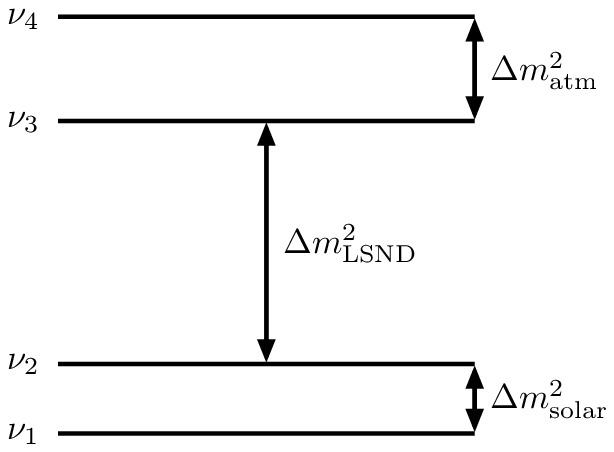}}
\vskip -8in \hspace{0.0cm}\caption[]{\small The type IIb four-neutrino
  mass pattern.  Interchange of the members in a given pair or the
  relative location of the two pairs may be allowed.}
\protect\label{fig1}
\end{figure}


\begin{figure}[!hb]
\vspace{-10pt}
\centerline{\hspace{-3.3mm}
\epsfxsize=8cm\epsfbox{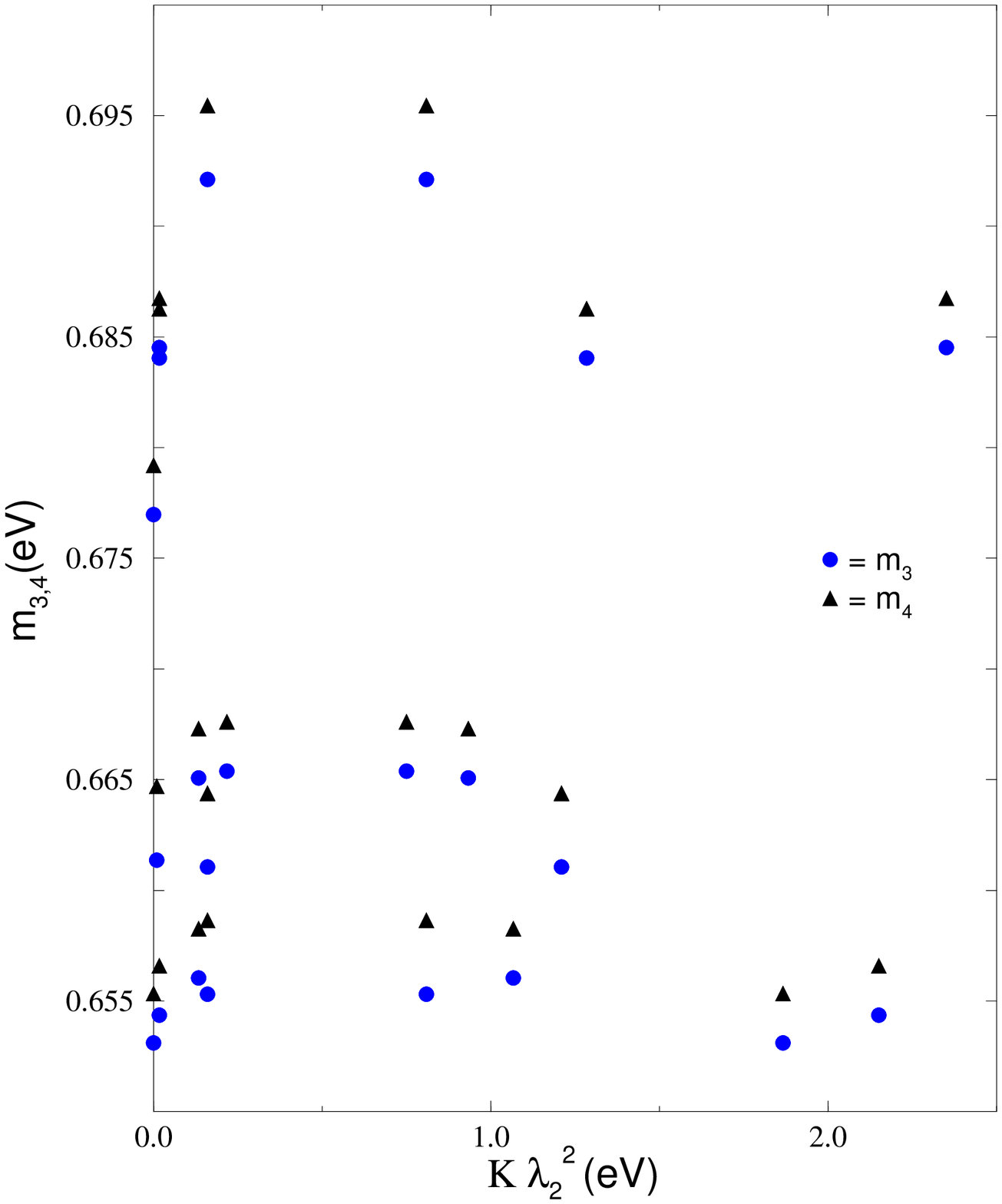}
\hspace{-0.1cm}
\epsfxsize=8cm\epsfbox{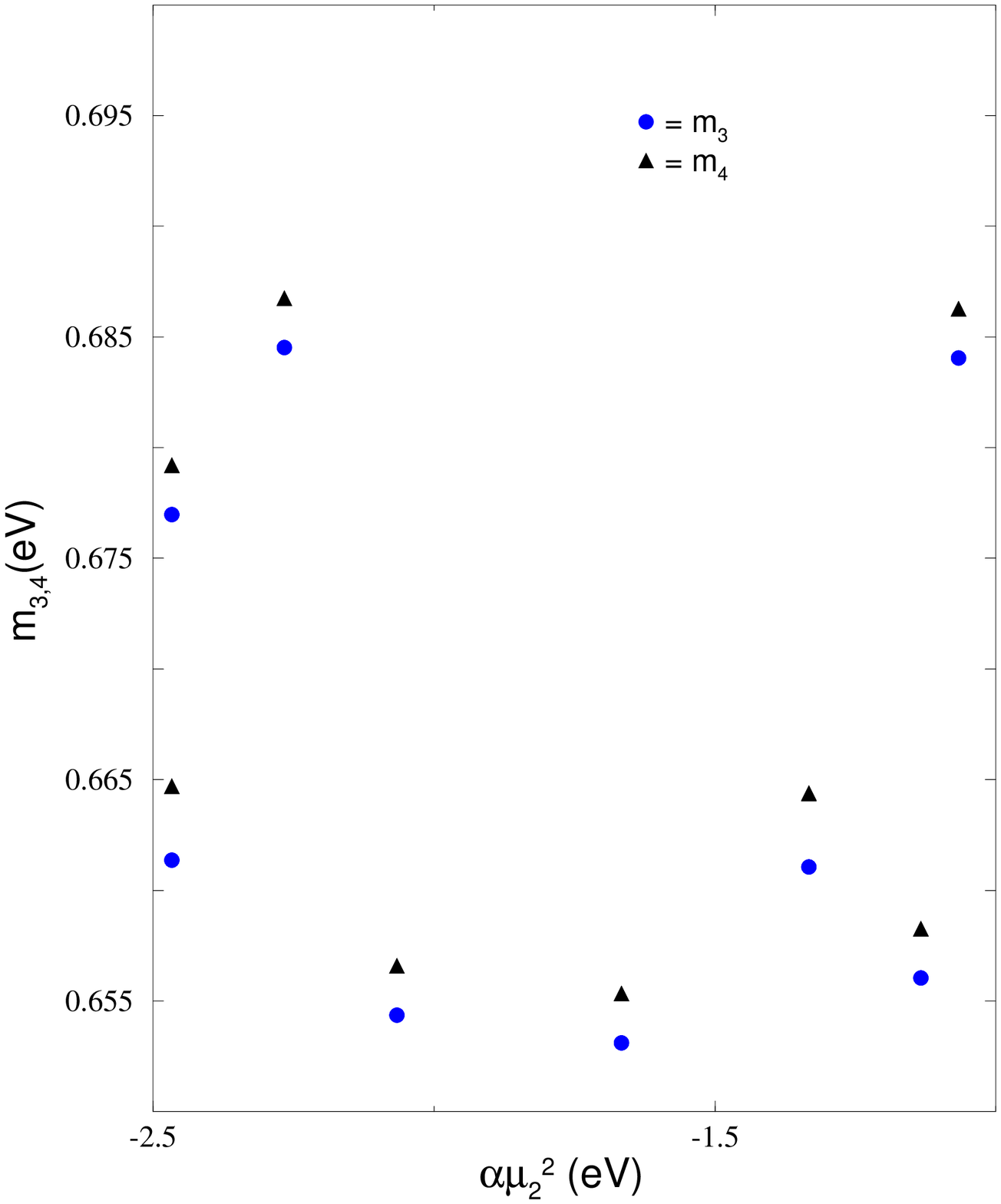}}
\centerline{\hspace{-0.5cm} (a) \hspace{7.0cm} (b)}
\hspace{3.3cm}\caption[]{\small Mass spectrum as a function of the 
(a) trilinear and (b) bilinear R-parity-violating parameters.}
\protect\label{fig2}
\end{figure}

\end{document}